\documentclass{iauc}
\usepackage{graphicx}
\usepackage{natbib}

\pubyear{2005}
\volume{199}
\pagerange{1--}
\jname{Probing Galaxies through Quasar Absorption Lines}
\editors{P. R. Williams, C. Shu, and B. M\'{e}nard, eds.}
\newcommand{\FeII}{\hbox{{\rm Fe}{\sc \,ii}}}

\newcommand{\MgI}{\hbox{{\rm Mg}{\sc \,i}}}
\newcommand{\MgII}{\hbox{{\rm Mg}{\sc \,ii}}}

\newcommand{\HI}{\hbox{{\rm H}{\sc \,i}}}

\newcommand{\msun}{M$_{\odot}$}

\newcommand{\mc}{\hbox{$\lambda 2852$}}
\newcommand{\fa}{\hbox{$\lambda 2600$}}
\newcommand{\LL}{\hbox{$\lambda\lambda$}}
\newcommand{\zab}{\hbox{$z_{\rm abs}$}}
\newcommand{\kms}{\hbox{${\rm km\,s}^{-1}$}}

\newcommand{\dNdz}{\hbox{${\mathrm d N}/{\mathrm d z}$}}
\newcommand{\drdz}{\hbox{${\mathrm d r}/{\mathrm d z}$}}

\newcommand{\nabs}{\hbox{716}}
\newcommand{\ngalsim}{\hbox{$\sim$~100,000}}
\newcommand{\ngalused}{\hbox{94,649}} 

\newcommand{\Arelcorrdrone}{\hbox{$0.67 \pm 0.09$}}
\newcommand{\LRGmassrangedrone}{$0.5$--$2.5\times 10^{12}$\,\msun}

\newcommand{\Arel}{\hbox{$0.87 \pm 0.08$}}
\newcommand{\Arelcorr}{\hbox{$0.69\pm 0.09$}}
\newcommand{\Arelcorrsys}{\hbox{$0.69 \pm 0.07\pm {0.06}$}}

\newcommand{\massratio}{\hbox{$7$--$10$}}
\newcommand{\LRGmassrange}{$1$--$2\times 10^{12}$\,\msun}
\newcommand{\overestimate}{\hbox{$25\pm10$}}
\title[Measuring the halo mass	of \MgII\ absorbers] %% give here short title %% 
{Measuring the halo mass
	of \MgII\ absorbers from their cross-correlation with Luminous Red Galaxies}

\author[N. Bouch\'e, M. T. Murphy, C. P\'eroux, \& I. Csabai]   %% give here short author list %%
{Nicolas  Bouch\'e$^1$, %\thanks{Email: nbouche@mpe.mpg.de},
M. T. Murphy$^2$, C. P\'eroux$^3$ \and I. Csabai$^4$}

\affiliation{$^1$Max Plank Institut f\"ur extraterrestrishe Physik,
Giessenbachstrasse, Garching D-85748, Germany; nbouche@mpe.mpg.de\\[\affilskip]
$^2$Institute of Astronomy, University of Cambridge, Madingley Road,
  Cambridge CB3 0HA, UK;\\[\affilskip]
$^3$European Southern
  Observatory, Karl-Schwarzschild-Str 2, D-85748 Garching, Germany;\\[\affilskip]
$^4$Dept. of Physics, E\"otv\"os Lor\'and University, Budapest, Pf. 32, H-1518
Budapest, Hungary.}

\begin{document}

\maketitle

\begin{abstract}
We study the cross-correlation between \nabs\ \MgII\ quasar absorption systems
and $\ngalsim$ Luminous Red Galaxies (LRGs) selected from the Sloan
Digital Sky Survey Data Release 3 in the redshift range
$0.4\!\leq\!z\!\leq\!0.8$. The \MgII\ systems were selected to have \LL
2796 \& 2803 rest-frame equivalent widths $\geq\!1.0$\,\AA\ and
identifications confirmed by the \FeII\ \fa\ or \MgI\ \mc\ lines. Over
co-moving scales 0.2--13$h^{-1}{\rm \,Mpc}$, the \MgII--LRG
cross-correlation has an amplitude \Arelcorr\ times that of the LRG--LRG
auto-correlation. Since LRGs have halo-masses of $10^{13}$\,\msun,
this strong cross-correlation signal implies that the absorber host-galaxies have halo-masses
\LRGmassrange. 
\keywords{cosmology: observations --- galaxies: evolution, halos ---
 quasars: absorption lines}
%% add here a maximum of 10 keywords, to be taken form the file <Keywords.txt>.

\end{abstract}

\firstsection % if your document starts with a section,
              % remove some space above using this command.
\section{Introduction}
The connection between quasar (QSO) absorption line (QAL) systems and galaxies
\citep[first established by][]{BergeronJ_91a} is important to our understanding of galaxy
evolution. QALs provide detailed information about the physical
conditions and kinematics of galaxies out to large impact parameters
($R\!>\!100{\rm \,kpc}$), regardless of the absorber's intrinsic
luminosity \citep[e.g.][and Churchill et al. 2005, Kacprzak et al. 2005, these proceedings]{SteidelC_02a}.
Past results show that \MgII\ absorbers are   biased towards late-type galaxies which do not evolve
strongly from $z\!\simeq\!1$ \citep{SteidelC_92a,SteidelC_94a}. These results also show that
\MgII\ absorber host-galaxies have $K$-band luminosities consistent with
normal $0.7L^*_B$ Sb galaxies. The cross-section of \MgII\
absorbers with $W_{\rm r}^{\rm MgII}\!\geq\!0.30$\,\AA\ appears to be 
$R_\times\!\sim\!70h^{-1}{\rm \,kpc}$ (co-moving)
\citep[e.g.][]{SteidelC_95b}. These systems are   associated
with \HI\ absorbers in the Lyman limit regime
  up to the damped Ly-alpha absorber (DLA) regime (see also Rao et al. 2005, these proceedings).

In \citet{BoucheN_04a}, we used the Sloan
Digital Sky Survey (SDSS) data release 1 \citep[DR1;][]{AbazajianK_03a}  to constrain  
the mass of the halos associated with the  \MgII\
absorbers. Specifically, we used the absorber-galaxy cross-correlation 
to measure of the mass ratio of the halos associated with \MgII\
since, in a hierarchical galaxy formation scenario, the amplitude ratio
of the \MgII--LRG cross-correlation to the LRG--LRG auto-correlation is also
their bias ratio.
The reader is referred to \citet{BoucheN_04a} and \citet{BoucheN_05b} for the details.
Fig.~1 (left) illustrates the methodology.
 Using 212 \MgII\ absorbers and
$\sim\!20,000$ Luminous Red Galaxies (LRGs), \citet{BoucheN_04a} found that the  bias
ratio $b_{\MgII}/b_{\rm LRG}$ is \Arelcorrdrone\   on scale $r_\theta\!>\!200h^{-1}{\rm \,kpc}$, implying a halo mass for
the \MgII\ host galaxies of 
\LRGmassrangedrone\ (for $10^{13}$\,\msun\  LRG halos).

%Throughout this paper, we adopt $\Omega_{\rm M}\!=\!0.3$,
%$\Omega_\Lambda\!=\!0.7$,  and $H_0\!=\!100 h\,\kms{\rm \,Mpc}^{-1}$. 

\section{Results}

\begin{figure}
\centerline{
 \includegraphics[width=52mm]{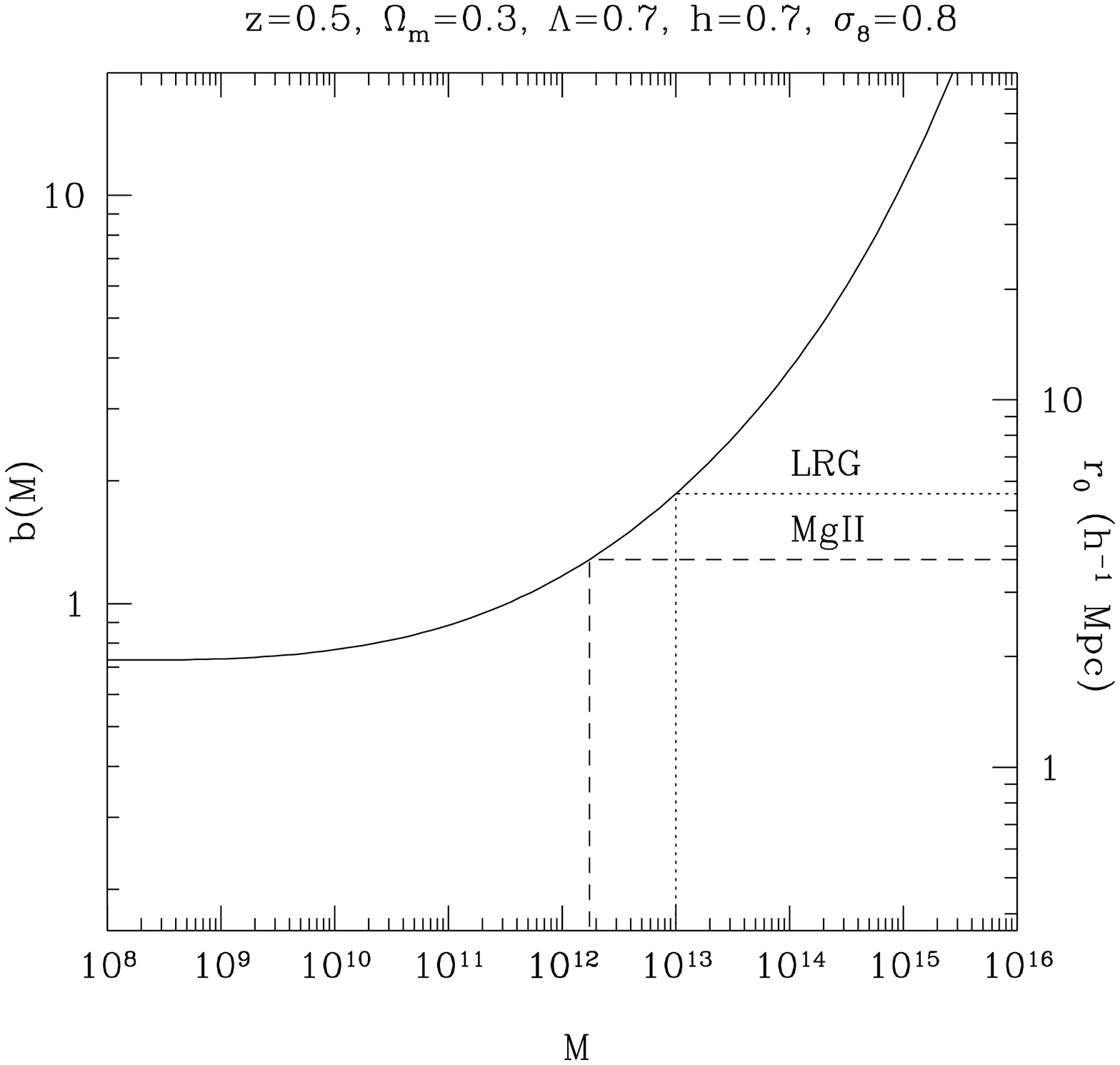}\hfill
 \includegraphics[width=52mm]{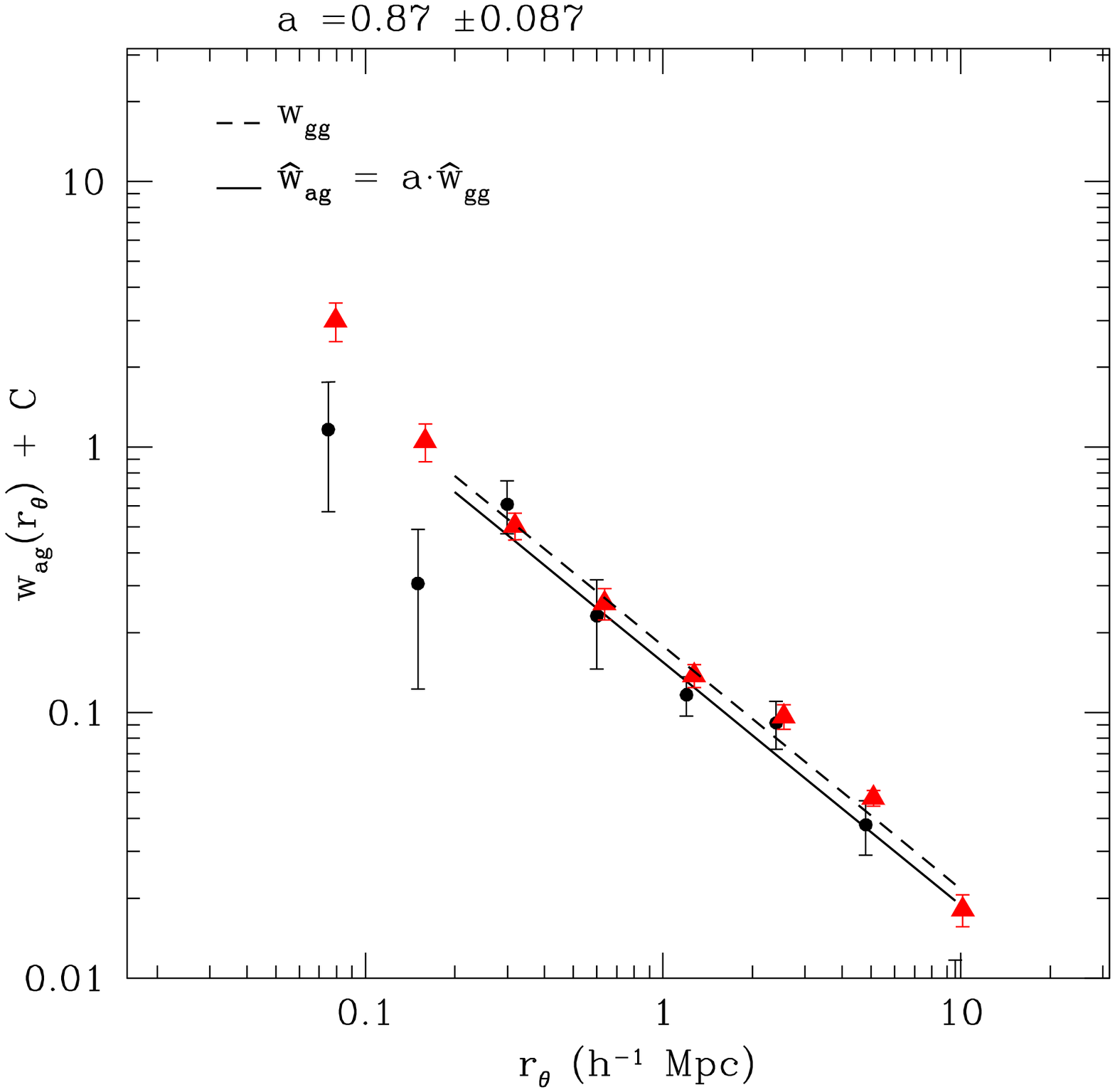}
}
\caption{{\it Left :}The bias $b(M)$ as a function of the halo mass
  $M$ \citet{MoH_02a} (solid line). The right $y$-axis shows the
  auto-correlation length $r_{\rm 0}$. LRGs have $r_{\rm
    0,gg}\simeq6~h^{-1}$~Mpc, and thus masses of
  $\simeq10^{13}$\,\msun. {\it Right}: Filled circles show the
  \MgII--LRG cross-correlation $w_{\rm ag}(r_\theta)$ between \nabs\
  \MgII\ absorbers and \ngalused\ LRGs.  Filled triangles show the
  LRG--LRG auto-correlation, $w_{\rm gg}$.  The dashed line shows a
  power-law fit to $w_{\rm gg}$.  The solid line shows the fit $\hat
  w_{\rm ag}=a\times \hat w_{\rm gg}$ for $r_\theta\!>\!200h^{-1}{\rm
    \,kpc}$ since the smallest scales will be affected by the finite
  cross-section of the absorbers.  The raw relative amplitude is
  $a\!=\!\Arel$. The left panel therefore implies that our \MgII\
  absorbers have halos 7--10 times less massive than LRG halos,
  i.e.~our \MgII\ absorbers have halos with mass
  \LRGmassrange.}\label{fig:result}
\end{figure}

Here, we extend our DR1 results using SDSS Data Release 3 \citep[DR3;][]{AbazajianK_05a}.
We selected \nabs\  \MgII\ absorbers	from SDSS/DR3 with $\zab \le
0.8$ using   automated technique that included the following criteria:
(i) $W_{\rm r}^{\rm MgII}\!\geq\!1.0$\,\AA\,;
(ii)    we require that  ${W_{\rm r}^{\rm MgI}\!\geq\!0.2}$\AA\,, and
that $W_{\rm r}^{\rm FeII}\!\geq\!0.5$ following the DLA criteria of
\citet{NestorD_03a} and \citet{RaoS_00a} (see Rao et al. 2005, these proceedings, for
an updated discussion).
We remove   spurious candidates by visually inspecting each \MgII\ spectrum.

For each absorber, we selected $\sim1300$ Luminous Red Galaxies (LRGs) from SDSS/DR3
using color criteria following
  \citet{ScrantonR_04a}, and in a slice of width
   $W_z=0.1$ using  the photometric redshift  calculated with the code of \citet{CsabaiI_03a}.
There are a total of \ngalused\ LRGs meeting these criteria, 
 within 12.8$h^{-1}$~Mpc, our largest bin.

For the cross-correlation, $w_{\rm ag}$, we use the   estimator
$1+ w_{\rm ag}(r_\theta)= {\rm AG}/{\rm AR}\,,$
where AG is the total observed number of absorber--galaxy pairs between
$r_\theta-\mathrm d r/2$ and $r_\theta+\mathrm d r/2$ and AR is the total
absorber--random galaxy pairs. This estimator is necessary to account for the non-symmetric
situation: \MgII\ absorbers have precise redshifts, while the LRGs have photometric redshifts
with an accuracy of $\sigma_z\;\simeq\;0.1$ \citep[see][for a discussion]{BoucheN_05b}.
 Fig.~\ref{fig:result}(right) shows
our results (see caption). The errors to $w_{\rm ag}$  and $w_{\rm gg}$ were
 computed using $N_{\rm jack}\!=\!10$ jackknife realizations.

The amplitude of the \MgII\--LRG cross-correlation
relative to that of the LRG--LRG auto-correlation is
 \Arelcorrsys, after applying a   correction of \overestimate\ per cent discussed   in
\citet{BoucheN_04a}.  
 The two error terms reflect  the statistical and systematic uncertainty,
respectively. By  adding the errors in quadrature, the bias ratio is
\begin{equation}
a=\Arelcorr\,.\label{res:arelcorr}
\end{equation}
Within the context of hierarchical galaxy formation,
Eq.~\ref{res:arelcorr} implies that our \MgII\ absorbers have  halo masses  \massratio\ times smaller
than the  LRGs.
For $10^{13}$\,\msun\ LRG halos, the \MgII\ absorbers have halos of \LRGmassrange.

It is important to realize that this method (i.e. measuring the halo mass from the ratio of projected correlation functions)
 has the following advantages \citep[see also][]{BoucheN_04a}:
(i)	it constrains the mass of the \MgII/DLA
	host-galaxies in a statistical manner without directly identifying them.
(ii)    it is free of systematics from contaminants (e.g.~stars),  
and (iii) it  does not require  knowledge of the true width of the redshift distribution of the galaxy used.
The last two points are a consequence of the fact that we use the same galaxies for $w_{\rm gg}(r_\theta)$ and for
	 $w_{\rm ag}(r_\theta)$.

\section{Discussion}
  
Our results are consistent with those of \citet{BergeronJ_91a},
\citet{MoH_96a}: For instance, \citet{MoH_96a} indicate that the
majority of \MgII\ absorbers reside in systems with
$V_{circ}=150$--$300$~\kms\ with a median at $\sim 200$~\kms. Our mass
measurement appears to corroborate that of \citet{SteidelC_94a} who
found that \MgII\ absorbers with $W_{\rm r}^{\rm
  MgII}\!\geq\!0.3$\,\AA\ are associated with late-type
$\sim\!0.7L_B^*$ galaxies, since the expected amplitude ratio between
early and late type galaxies is $\sim\!0.70$
\citep[see][]{BoucheN_04a}.

Are our results consistent with $\Lambda$CDM? That is, is there enough
massive halos to account for $\dNdz$?  From
$\dNdz=n(M)\;\sigma(M)\;\drdz,$ $R_X\simeq70$~kpc (co-moving)
\citep{SteidelC_95b}, $n(M)=10^{-2}h^{-3}$~Mpc$^{-3}$,
$\dNdz=0.3\;(n/10^{-2})\;(R_X/70~\hbox{kpc})^2\simeq
\dNdz(\hbox{obs}),$ and we can conclude that there are enough massive
$10^{12}$~\msun\ halos.
While we defer a more detailed analysis of these results to \citet{BoucheN_05d},  
preliminary results also indicate little dependence of the halo mass on the
equivalent width.

\begin{acknowledgments}
  N.B.~thanks the IAU and the organizers of this conference for the
  travel grant.  M.T.M.~thanks PPARC and I.C.~thanks OTKA, T047244.
  Funding for the Sloan Digital Sky Survey (SDSS) has been provided by
  the Alfred P. Sloan Foundation, the Participating Institutions, the
  National Aeronautics and Space Administration, the National Science
  Foundation, the U.S.  Department of Energy, the Japanese
  Monbukagakusho, and the Max Planck Society.

\end{acknowledgments}

%\bibliography{references}

\begin{thebibliography}{}
\small
\itemindent -0.48cm

\bibitem[\protect\citeauthoryear{{Abazajian}~{et al.}}{{Abazajian}
  {et~al.}}{2003}]{AbazajianK_03a}
{Abazajian} K.~{et~al.,} 2003, AJ, 126, 2081

\bibitem[\protect\citeauthoryear{{Abazajian}~{et al.}}{{Abazajian}
 {et~al.}}{2005}]{AbazajianK_05a}
{Abazajian} K.~{et~al.,} 2005, AJ, 129, 1755

\bibitem[\protect\citeauthoryear{{Bergeron} \& {Boiss{\'e}}}{{Bergeron} \&
  {Boiss{\'e}}}{1991}]{BergeronJ_91a}
{Bergeron} J.,  {Boiss{\'e}} P.,  1991, A\&A, 243, 344

\bibitem[\protect\citeauthoryear{{Bouch{\' e}}, {Murphy} \& {P{\'
  e}roux}}{{Bouch{\' e}} et~al.}{2004}]{BoucheN_04a}
{Bouch{\' e}} N.,  {Murphy} M.~T.,    {P{\' e}roux} C.,  2004, MNRAS, 354, 25L

\bibitem[\protect\citeauthoryear{{Bouch{\'e}}~et al.}{{Bouch{\'e}}~{et~al.}}{2005a}]{BoucheN_05b}
{Bouch{\'e}} N.~{et al.}, 2005a, ApJ, in press, astro-ph/0504172

\bibitem[\protect\citeauthoryear{{Bouch{\' e}}, {Murphy}, {P{\' e}roux} \&
  {Csabai}}{{Bouch{\' e}} et~al.}{2005b}]{BoucheN_05d}
{Bouch{\' e}} N.,  {Murphy} M.~T.,  {P{\' e}roux} C.,    {Csabai} I.,  2005b,
  MNRAS, in prep.


\bibitem[\protect\citeauthoryear{{Csabai}~{et al.}}{{Csabai}~{et al.}}{2003}]
  {CsabaiI_03a}
{Csabai} I.~{et~al.,} 2003, AJ, 125, 580

\bibitem[\protect\citeauthoryear{{Mo} \& {Miralda-Escud{\'e}}}{{Mo} \&
  {Miralda-Escud{\'e}}}{1996}]{MoH_96a}
{Mo} H.~J.,  {Miralda-Escud{\'e}} J.,  1996, ApJ, 469, 589

\bibitem[\protect\citeauthoryear{{Mo} \& {White}}{{Mo} \&
  {White}}{2002}]{MoH_02a}
{Mo} H.~J.,  {White} S.~D.~M.,  2002, MNRAS, 336, 112

\bibitem[\protect\citeauthoryear{{Nestor}, {Rao}, {Turnshek} \& {Vanden
  Berk}}{{Nestor} et~al.}{2003}]{NestorD_03a}
{Nestor} D.~B.,  {Rao} S.~M.,  {Turnshek} D.~A.,    {Vanden Berk} D.,  2003,
  ApJ, 595, L5

\bibitem[\protect\citeauthoryear{{Rao} \& {Turnshek}}{{Rao} \&
  {Turnshek}}{2000}]{RaoS_00a}
{Rao} S.~M.,  {Turnshek} D.~A.,  2000, ApJS, 130, 1

\bibitem[\protect\citeauthoryear{{Scranton}~{et al.}}{{Scranton}~{et al.}}{2003}]
  {ScrantonR_04a}
{Scranton} R.~{et~al.,} 2003, Phys.~Rev.~Lett., submitted, preprint (astro-ph/0307335)


\bibitem[\protect\citeauthoryear{{Steidel} \& {Sargent}}{{Steidel} \&
  {Sargent}}{1992}]{SteidelC_92a}
{Steidel} C.~C.,  {Sargent} W.~L.~W.,  1992, ApJS, 80, 1


\bibitem[\protect\citeauthoryear{{Steidel}, {Dickinson} \& {Persson}}{{Steidel}
  et~al.}{1994}]{SteidelC_94a}
{Steidel} C.~C.,  {Dickinson} M.,    {Persson} S.~E.,  1994, ApJ, 437, L75

\bibitem[\protect\citeauthoryear{{Steidel}}{{Steidel}}{1995}]{SteidelC_95b}
{Steidel} C.~C.,  1995, in {Meylan} G.,  ed., QSO Absorption Lines.
  Springer-Verlag, Berlin, Germany, p.~139


\bibitem[\protect\citeauthoryear{{Steidel}, {Kollmeier} {et~al.}}{{Steidel} et~al.}{2002}]{SteidelC_02a}
{Steidel} C.~C.,  {Kollmeier} J.~A. et~al.,  2002, ApJ, 570, 526


\end{thebibliography}
%\bibliographystyle{mn2e}

%\begin{discussion}

%\discuss{Massey}{I'm wondering if you have considered the expected intrinsic 
%dispersion in absolute magnitude of WRs -- if you consider the (large) mass 
%range that becomes an early WN or late WC according to the evolutionary models, 
%wouldn't you expect a large dispersion in M$_v$?}

%\discuss{van der Hucht}{Indeed, we will be always left with some intrinsic 
%scatter in M$_v$ due to mass differences within the same spectral subtype. But 
%in my opinion, the current large dispersion is for a large fraction due to 
%incertainties of the adopted distances of open clusters and OB associations.}

%\end{discussion}

\end{document}